\documentclass[journal,comsoc]{IEEEtran}
\usepackage[T1]{fontenc}
\usepackage[utf8]{inputenc}

\usepackage{cite}

\usepackage{graphicx}
\usepackage[justification=centering]{caption}

\usepackage{amsmath}
\usepackage[cmintegrals]{newtxmath}

\begin{document}

\title{Privacy-preserving classifiers recognize shared mobility behaviours from WiFi network imperfect data}

\author{

\IEEEauthorblockA{Orestes~Manzanilla-Salazar\IEEEauthorrefmark{1} and
Brunilde~Sans\`o\IEEEauthorrefmark{2}}

\IEEEauthorblockA{
Email: \IEEEauthorrefmark{1}orestes.manzanilla@polymtl.ca,
\IEEEauthorrefmark{2}brunilde.sanso@polymtl.ca}

\thanks{O. Manzanilla-Salazar is with the Department of Electrical Engineering of Polytechnique Montr\'eal, QC, Canada,  (e-mail: orestes.manzanilla@polymtl.ca) }
\thanks{B. Sans\`o is with the Department of Electrical Engineering of Polytechnique Montr\'eal, QC, Canada,  (e-mail: brunilde.sanso@polymtl.ca) }
\thanks{This work has been submitted to the IEEE for possible publication. Copyright may be transferred without notice, after which this version may no longer be accessible.}

}


\maketitle

\begin{abstract}

This paper \emph{proves the concept} that it is feasible to accurately recognize specific human mobility shared patterns, based solely on the connection logs between portable devices and WiFi Access Points (APs), while preserving user's privacy. We gathered data from the Eduroam WiFi network of Polytechnique Montreal, making omission of device tracking or physical layer data. The behaviors we chose to detect were the movements associated to the end of an academic class, and the patterns related to the small break periods between classes.

Stringent conditions were self-imposed in our experiments. The data is known to have errors  noise, and be susceptible to information loss. No countermeasures were adopted to mitigate any of these issues. Data pre-processing consists of basic statistics that were used in aggregating the data in time intervals.

We obtained accuracy values of 93.7 \% and 83.3 \% (via Bagged Trees) when recognizing behaviour patterns of breaks between classes and end-of-classes, respectively.

\end{abstract}


\IEEEpeerreviewmaketitle

\section{Introduction}

\IEEEPARstart{P}{ublic} indoors facilities like hospitals, universities, airports and malls often offer WiFi access services to visitors and employees. One example of such networks is the "Eduroam" network service offered in many academic institutions worldwide providing Internet access to faculty, employees and students \cite{griffioen17}.   

People leave \emph{footprints} in the data generated by network administrative systems and in the portable devices being carried with them as they move through the space. We classify these footprints as:

\begin{figure}[!t] 
	\centering
	\includegraphics[width=2.7in]{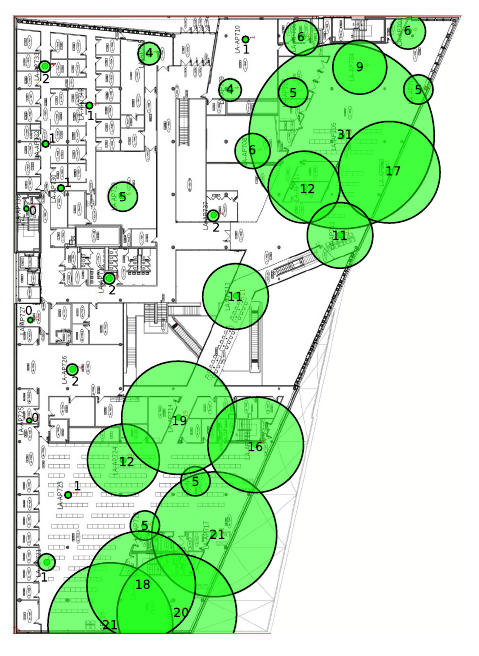}
	\centering	
	\caption{Count of devices connected to each AP in one floor of the building.}
	\label{lassondeplan}
\end{figure}

\begin{itemize}
	\item \emph{Device-free network-gathered:} information about individuals is obtained whether they carry a device connected to the network or not, by observing the perturbations caused in the wireless signals. People's bodies absorve or reflect the electromagnetic waves affecting measured values of the physical layer (e.g.: signal strength, phase, etc.) \cite{li18} \cite{wang15b} \cite{ali15}.
	\item \emph{Device-gathered:} this refers to changes registered in the client devices connected to the network, either by active sensors (GPS, accelerometers, compass, light sensors) or connection data (e.g.: time of flight, signal strength, etc.)  \cite{mun08} \cite{kjaergaard12} \cite{manweiler13}.
	\item \emph{Device-enabled network-gathered:} these are changes registered by fixed APs or Base Station (BS) to which a device (moving node) is connected. This encompasses both connection characteristics (e.g.: time of flight, signal strength, etc.) as well as the status of connection between a device and a particular AP or BS. 
\end{itemize}

\begin{figure}[!t] 
	\centering
	\includegraphics[width=2.7in]{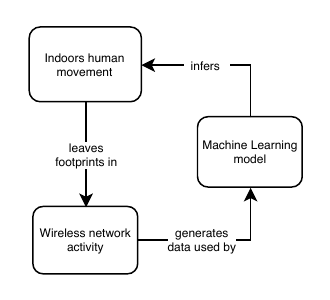}
	\centering	
	\caption{Subsystems.}
	\label{subsystems}
\end{figure} 

This paper deals with the third type of footprint. Our objective of this work is to do a \emph{proof of concept} that ML methods can be applied to detect human behaviour under stringent data gathering conditions. In particular, even though we make inference from footprints of the third class, we exclusively use the \emph{Eduroam} logs of the devices connected to each AP, in two floors of one building of Polytechnique Montreal in Canada (Fig. \ref{lassondeplan}). We do not use physical layer measurements of any kind, nor do we track the user. We show that it is possible to push the limits of both the austerity of the data used, and the simplicity of the inference techniques, while maintaining the ability to infer activities related to the movement of people. In this paper, the activities chosen are the movement patterns related to the moment in which a class has finished, and those related to periods of time when students have a break between classes.  

We identify three systems (Fig. \ref{subsystems}):

\begin{itemize}
	\item \emph{Indoors human movement:} this refers to the humans transiting the space under study, their activities, in particular those implying movement big enough to trigger a hand-over in a device connected to the WiFi network, as is done in \cite{poucin18}. As mentioned above, in our case, the behaviours of interest in this system are circumscribed to the way academic activities are scheduled in the university. Another element that gives structure to the trajectories followed by people within the building, is its physical structure and layout.
		\item \emph{Wireless network:} this encompasses the set of APs in the floors considered in the data gathering process, the set of portable or mobile devices carried by people within the building, connected to the network, and the components of the Cisco Connected Mobile Experiences (CMX) solution, which is used by the wireless network administrator in Polytechnique Montreal.
		\item \emph{Machine learning models:} this is  constituted by the models of supervised learning used to infer activities in the human system from data of the wireless nework system. The aim of learning about human activities has been proved to be feasible in small scale activities like rising an arm \cite{li18}, larger trajectories \cite{gonzalez08} \cite{song10b}, activities involving groups of people in large spaces \cite{kjaergaard12} \cite{haghani17} and routines \cite{eagle09}.

\end{itemize}
 
Why is it important to infer patterns of human behavior from wireless data? Because when people's health \cite{tan18}, integrity and security is put at risk \cite{song17}, having the ability to \emph{predict}, \emph{detect} and \emph{understand} human behavior is useful to have timely reactions, and more precise mitigation strategies. There are other uses for which detecting human behavior patterns is useful, like crowdsensing, recommendation systems, social networks \cite{guo17}, as well as classification and understanding of the use of physical spaces \cite{poucin18} \cite{caceres18} \cite{ruiz14} \cite{toole12} \cite{calabrese10}.

We show in this paper a new strategy to study human activity via patterns in WiFi AP logs  \cite{mun08} \cite{kim05}. Instead of focusing in differenciating spaces by how they are used \cite{poucin18}, or on individual mobility, our goal is to detect of specific patterns of simultaneous behaviour in groups of people. The approach takes into consideration the following premises:

\begin{enumerate}
	\item The data required should be easy to obtain in widespread WiFi networks.
	\item Pre-processing and machine learning techniques as well as storage requirements should be as simple and cheap as possible.
	\item Both the data required for the techniques proposed and the information that is inferred from it, should respect users privacy.
\end{enumerate}

In order to comply with our premises, we intentionally refrain from:
\begin{itemize}
	\item Analysing individual data.
	\item Applying filters to eliminate redundant data.
	\item Applying filtering to eliminate data from individuals that do not follow the specific pattern that is being detected (e.g.: not following the academic hourly schedule).
	\item Implementing methods to identify and correct errors in the data.
	\item Applying techniques to recognize and repair cases of incomplete data.
	\item Using any context information from the infrastructure or the date and time of a measurement, when using a trained machine learning model to detect a pattern.
	\item Using wireless Channel State Information (CSI).
\end{itemize}

We show that notwithstanding the stringent conditions self-imposed, it is feasible to detect specific simultaneous behaviour patterns, such as the end of a class of an academic course, or that people are in one of the small breaks between classes. To the best of our knowledge this is the first time that movement patterns have been successfully recognized based on aggregated  device counts per AP in a WiFi network, without CSI or physical layer values. 

The rest of the article is comprised by a review of the related literature, a section on the issues and challenged involved, and a detailed description of the  three systems mentioned: the human behaviour, the WiFi network, and the machine learning models used to infer human behaviour (Fig. \ref{subsystems}). We will then present our results and analysis in the following section and will end the article discussing our conclusions and future work.

\section{Related work}

In order to put our work into context, we will briefly mention the main approaches that can be given to the study of spatio-temporal patterns in human behaviour, as well as the ways in which data has been gathered in other empirical studies. Lastly we will mention some of the ways in which this kind of data has been analysed in previous works.

\subsection{Spatio-temporal patterns in human behavior}

Spatial information regarding human activity can be classified as \emph{pose} \cite{toshev14} and micro-activities \cite{li18} \cite{tan18} \cite{wang18b}, meaning the change of the shape of the space the human body occupies at a particular moment, and \emph{trajectories} or the history of location through time. Human trajectories have enough regularity as to make it feasible to look for patterns in them \cite{gonzalez08} \cite{song10b}. Also, human routines have enough structure as to allow the analysis of patterns \cite{eagle09}.

The focus of the analysis of human movement can be placed on individual patterns (generally trajectories or tracking data), or in groups of people (crowds or flocks), who exhibit a similar simultaneous behavior.

\subsubsection{Trajectory analysis}
The study of patterns in trajectories \cite{gonzalez08} \cite{song10b} requires analyzing a time-serie of location coordinates or indicators of some kind. Recently deep learning techniques have been used to represent trajectory patterns \cite{shah16} \cite{wu17} \cite{alahi16}. Location represents a great concern in terms of privacy, which has motivated researchers to create algorithms to anonymize location data \cite{wang18}. There is however a public concern of the possibility that an individual's anonymized data can be used to reconstruct the identity of the person (de-anonymization) \cite{al18}.

\subsubsection{Crowd analysis}
Crowd Analysis has gained attention as stampedes in places with high density of people turns them into an additional risk for themselves, besides the threat that can be assumed to originate the stampede. Some work has been done in the prediction of the formation of crowds \cite{huang18}. A recent and thorough review on the empirical studies of crowds can be found in \cite{haghani17}. 

\subsubsection{Data gathering methods}
Data from records of people activities, surveys, interviews, video surveillance, as well as from experiments with human volunteers, evacuation drills, natural disasters, virtual reality, and animals \cite{motsch17} \cite{haghani17} have helped build models to understand and predict the behavior of crowds, in addition to networks data. We group the sources of data in the following categories:

\begin{itemize}

	\item \emph{Video-based:} crowd behavior has been analyzed by extracting patterns from videos \cite{grant17}. Machine learning techniques allow the processing of videos from one or multiple cameras to identify both individual and crowd movement \cite{shao15}, as well as perform pose estimation \cite{toshev14}, and recognition of frames where relevant events take place \cite{gan15}. Deep learning techniques can recognize a posture simultaneously captured from different viewpoints \cite{rahmani15}, and in people identification based on range sensors such as LiDAR and RGBD cameras, which provide data in the form of 3D point clouds \cite{haque16}. One of the main issues with video is that lighting, obstruction, and perspective can affect the quality of the analysis. 3D point clouds are not affected by light conditions, but the positioning and obstacles can still represent a challenge, besides being a more expensive technology.
  
	\item \emph{Network-based:} GPS traces are one of the most obvious sources of location data. Traces from taxi drivers have been used to study patterns in their mobility. This kind of traces, however, are of limited application in indoors and urban environments \cite{aschenbruck11}, as obstacles can produce a shadowing effect. Devices GPS traces or data from GTFS networks can be used along with additional information, like WiFi traces \cite{zahabi17}, to overcome these issues. Applications installed on the client side can also provide information in the form of surveys like DataMobile \cite{patterson16}, which collects travel information. Gathering users data from their devices, however, presents the limitation of requiring users to share information regarding their position and trajectories, which is frequently avoided by users \cite{su04}. This is where monitoring centered in the network (fixed nodes) overcomes this problem as the information gathering becomes a passive and automatic process regularly carried out by Wi-Fi or GSM networks\cite{nguyen07}. This kind of data gathering presents low costs, and the pervasive nature of wireless networks opens the possibilities of collecting large quantities of data easily available in the administration software of public places that offer wireless connection \cite{calabrese13}. One of the problems to be overcomed with network data, is the low precision in location estimations \cite{mao07} \cite{wymeersch09}, though it can be improved by considering it along with other types of information \cite{aschenbruck11}. This motivates, in some studies, the use of symbolic spaces instead of geographical coordinates \cite{meneses12}. One example of a symbolic location is the identification of the AP to which a devices is connected. Some work is required in order to find the relationship between connections to APs and specific physical spaces \cite{poucin18}. WiFi and Bluetooth connections between user's devices can give information about the social network among the people present in a particular area, which can complement information about their behavior. Developments in multi-hop networks \cite{conti07} have brought increasing attention in this kind of data. Studying movement within a WiFi networks via connection logs of connections established with APs,  indirectly provides low cost large scale information about realistic natural trajectories. In \cite{poucin18} the same raw data we use in our research is used to infer the kind of use that is given to the space where each AP is. Nine categories of places are found via clustering. The focus is that they find patterns in places, whereas we find patterns in time. Because of this fact, the information is aggregated per day, instead of per minute.

	\item \emph{sensorless:} the effects the presence of human bodies have over signal strength make it feasible to  estimate the number of people in an area \cite{yoshida15}, without requiring every person in it to carry a device. Also signal strength combined with phase alterations, time-of-flight, among other signal characteristics allow the detection of activities \cite{toshev14} and micro-activities \cite{li18} \cite{tan18} \cite{wang18b}.
	
	\item \emph{sensor-based:} some research has been done to identify the human activity using sensors data, like signal strength (the use of the wireless signal as a sensor), device accelerometer, light sensors, compass, etc \cite{manweiler13}. Our research goes in the opposite direction of these methods, for our premise is to take advantage of simple and widespread existing technology, instead of working with additional and more sophisticated data gathering equipments and techniques.
	
\end{itemize}

\subsection{Network Analysis}
Various data analysis and inference techniques have been proposed and used in order to extract spatio-temporal patterns in the behaviour of people, from networks data.

In the analysis of WiFi data, looking at changes in number of connections to a network's APs from the point of view of the frequency domain allows the identification of the most important periodicities in data \cite{kim07}. 

Besides the techniques that have been mentioned in the modelling and classification of traces, there are another tasks of relevance. There is the ability to predict tasks patterns related specifically to individuals in a particular AP, as is shown in \cite{manweiler13} where length of stay of a device in a particular AP is predicted based on Wi-fi data along with information from other sensors in the device. We classify the focus of the analysis of the network traces in three major categories:
  
\begin{enumerate}
	\item \emph{Focused on spaces:} the analysis of mobility patterns can be aimed at studying how  spaces are used \cite{poucin18} \cite{caceres18} \cite{ruiz14} \cite{toole12} \cite{calabrese10}. The subject over which information is  being inferred are physical spaces.
	\item \emph{Focused on individuals:} on the other hand, the subjects over which inference is applied, can be the individuals moving through space. In \cite{mun08} a technique is proposed to classify individual behavior based on the hand-over patterns of users' devices among various antennas in a GSM or Wi-Fi network. In \cite{kim05} wireless traffic data is used to build models of individual users' mobility. A combination of focus on both individuals and spaces is observed in tasks like that proposed in  \cite{manweiler13} (predict length of stay at a particular individual connected to a specific AP).
	\item \emph{Focused on groups:} here the focus is understanding movement related to specific behavior shared by groups, flocks or crowds. The analysis of WiFi logs, for example, permits the identification of groups of individuals moving ``together'', or flocks, within indoor environments \cite{kjaergaard12}. We position our research in this category,  as we take advantage on the fact that the aggregation of individual data, which is beneficial from the point of view of privacy, permits the detection of shared behaviours.
\end{enumerate}

\section{Issues and challenges}

We now bring to attention the two main issues inherent to the nature of our research and its premises:

\subsection{Privacy issues} 
One of the main challenges in the use of machine learning and statistical inference to analyze people's data is the need for a balance between the protection of privacy and innovation \cite{horvitz15}. In the case of AP connection logs, users' MAC addresses, the devices model identification, as well as the user's network identification are sensitive information, which brings the necessity of implementing some kind of anonymization technique. Not only this eliminates the collection of socio-demographic information \cite{poucin18}  but also prohibits using knowledge about the roles of a person in an area of study. Some researchers reconstruct the device information (distinguishing between mobile devices and laptops) \cite{poucin18}, users' roles (in an academic setting the roles could be those of students, faculty, staff, visitors, etc.) \cite{ruiz14} or even the identity of the users \cite{srivatsa12}. For some groups, however, the inference of anonymized data should be avoided and even regulated for privacy reasons \cite{horvitz15}. More details on de-anonymization can be found in \cite{ding10b}. Refraining from using sensible data, as well as reconstructing (de-anonymizing it) it using ML, is the most conservative approach.

\subsection{Quality of information challenges} 
The quality of the APs logs data is dependent on the following:
\begin{itemize}

	\item The status of the mobile or portable device WiFi activation. Users have the choice of turning off the WiFi antenna of their devices, or the whole device can be turned off, which makes the device invisible to the network. We consider this to be a problem of \emph{incompleteness of the data}.
	\item Device specific behaviour can vary depending on the battery level, and other settings, affecting the time that the WiFi passes without transmitting to the network. During this time, the device is also invisible to the network, and thus another contribution to \emph{incompleteness of the data}.
	\item Because of the limited range of connection to the APs, devices can get out of range, becoming also invisible to the network. Recognizing a device  coming back from a departure from the network as the same device, in the context of hashed MAC addresses, requires maintaining in memory a record of all the hashed generated. If the hashes are randomly generated each time devices connect again to the network, adding these new logs to the previous history of the device becomes a challenging task. We consider giving a different identification code to a same device in different moments, as \emph{a source of errors} in the data.
	\item The hand-over of the connection of a device to the network, from one AP to another, is triggered in general by problems in the quality of the connection, and availability of a better AP to connect with. Interpreting the hand-over as an event caused by the movement of the device, will introduce errors in the data, as it can also be caused by obstructions or by the so-called Ping-Pong effect that arises when the connection of device is quickly handed-over between two or more APs that are within range, even though the device might be geographically static \cite{poucin18}. We consider this to be \emph{a source of noise} in the data.
	\item The location of the AP with which a device has established a connection might not be the one physically closest to it. Therefore, symbolically assigning the location of the AP with which a connection is established, as an approximation of the general location of a device might be inaccurate. In fact, a connection might be established with an AP that is in a different floor, or a different room in the building. This phenomena can also be considered as \emph{a source of noise} in the data.
	 
\end{itemize}

We decided to accept the aforementioned issues and challenges as constraints. In the following section we go into detail about the way we approach the human activities, the network, and the use of pre-processing and ML to take advantage of the data in spite of the limitations and premises.

\section{Models Proposed}

We introduce now models of the three systems we have defined in our research (Fig. \ref{subsystems}):

\subsection{Modelling of human behavior}

The main elements of the human system are the people transiting the two floors considered for the study. These are composed by students, faculty, administrative staff and employees, and visitors.  Students, faculty, and occasionally members of administrative staff and employees, are considered to have their activities affected in various levels, by the weekly schedule of classes. 

Students enrol in courses whose classes take place within pre-established academic \emph{blocks}. These blocks start at 8:30 am, 9:30 am, 10:30 am, 11:30 am in the morning of work days, and have a duration of 50 minutes, leaving an interval of 10 minutes that can be used either as a break between two hours of a same course, or to go to the next class.  Afternoon blocks start at 12:45 pm, and then on each hour until 4:45 pm, with the same duration. We  consider only data from 8:30:00 am to 11:29:59 am, as most classes take place during the morning hours. Also, we exclude weekends, as no regular classes are scheduled on them. 

In our analysis, we discard data from Wednesdays, as these are days of the week in which a great proportion of the faculty has departmental meetings in Polytechnique Montreal.

The human dynamics of interest in this system, are those behaviours that, being shared by ideally large quantities of people, are associated to activities implying movements through the indoors spaces, large enough to trigger a hand-over of the connection of a mobile or portable device. We focus in shared behaviours that can be easily labelled, for being ruled by the weekly academic schedule.

The term "rule", however, could be too strong, as in reality it is suggested for the faculty and instructors whose classes encompass two or more blocks, to leave the 10 minutes interval between the end of a block and the start of the following one for students to take a break, but such suggestions can be ignored by some faculty or instructors. In some courses the break can be postponed for the sake of the continuity of the teaching process. When the 10-minutes interval occur between two different courses, students use this time to attend to personal needs and walk to the next class. There is no guarantee, however, that after the start of a break, or the end of a class, students will actually leave a classroom, nor that students will arrive exactly at the beginning of a block.

We defined two behaviours of interest:
\begin{itemize}

	\item The intervals of break between two blocks of class, and
	\item The moment in which a class finishes.
	
\end{itemize}

The first one will allow us to associate data from a particular moment to a behaviour that is defined as having some duration. The second will allow the detection of a particular event in time (theoretically instantaneous).

Alongside  the presence of students and faculty following the schedule in a particular moment, we assume there will be presence of human activity not ruled by this academic schedule as administrative staff, employees, visitors and students who are not attending to class may wander about in halls, stairs, bathrooms, food courts, as well as unoccupied classrooms, which will be also within the range of the APs.

\subsection{Modelling of the network}

The Eduroam WiFi network in Polytechnique Montreal, is monitored via the Cisco CMX solution, which constitutes our main instrument for data gathering. A script was written, to refresh the web interface that samples information from the devices connected to the APs of the network. One month of data, totalling 189,259 samples was gathered, in sampling intervals averaging 12 seconds. Samplings are not regularly spaced in time because each time the refresh is requested, the system needs to receive data from all the APs, which imposes a variable delay. Each sample provided a table with all the connections active between devices and APs in the area under study. A total of 6,830,873 connection logs between devices and APs were retrieved.

A simple random hashing process was used to anonymize the addresses, device models and users identifications for each device. When a particular device receives a hash, it is remembered during the period over which the device maintains its connection. When it disconnects, the hash is "forgotten". When and if the device connects again to the network, a new random hash will be generated. 

\begin{table}[]
	\centering	
	\caption{Typical raw data sampled from AP logs.}
	\label{sample}
\begin{tabular}{clll}
Sample ID & UserName & AP MAC Address & Sample time-stamp \\
1 & 5bac0b... & e0ed722608... & 2017-04-07 15:51:07 \\
1 & d704b5... & e934a10b73... & 2017-04-07 15:51:07 \\
2 & f059ea... & 5b50722d57... & 2017-04-07 15:51:52 \end{tabular}
\end{table}

We do not treat the data to fix issues like noise, errors and incompleteness. In particular, we refrain from the following:
\begin{itemize}
	\item Implementing any method to differentiate between types of devices (un-anonymization is against our premises).
	\item Implementing any method to recognize when two devices belong to the same user.
	\item Detect Ping-Pong sequences of hand-overs of a device.
	\item Implementing any method to recognize when a device connected in different times with different random hashes is, in fact, the same device.
	\item Using location estimates nor signal strength, that are given by the Cisco CMX solution in each sample of a connection between an AP and a device.
	\item Inferring when a disconnected device has left the area of study, or not.
	\item Inferring when a hand-over has been triggered by movement, or not.
	\item Estimating the geographical location of a device.
	\item Including information regarding the status of the APs.
\end{itemize}

Three examples of the raw data gathered are shown in Table \ref{sample}. The first two rows correspond to two devices connected to different APs, reported during the sampling number 1, requested to the administrative software at 15:51:07. The third one is a different device from the sampling number 2, that takes place some seconds later. 
We emphasize that when a new sampling is requested, both its consecutive ID and time-stamp will be associated to \emph{every row} of data generated in that request. For each active connection at that instant between a device and an AP, one row will be generated, sharing this information, as is show in table \ref{sample}. The time-stamps will be used in order to aggregate the rows of data.

\subsection{Modelling the learning process}

The problem is posed as one of supervised learning problem where each input instance corresponds to a vector of statistics related to APs and the output or label for the interval is manually assigned for each of the experiments.

The input, or feature vector $x_i$, is formed by pre-processing the data provided by the network administration software to calculate the minimum, maximum, variance and average of the number of devices connected to each one of the 67 APs distributed in two floors of a building of  Polytechnique Montr\'eal (Fig. \ref{lassondeplan}). 

\begin{figure}[!t]
	\centering
	\includegraphics[width=2.7in]{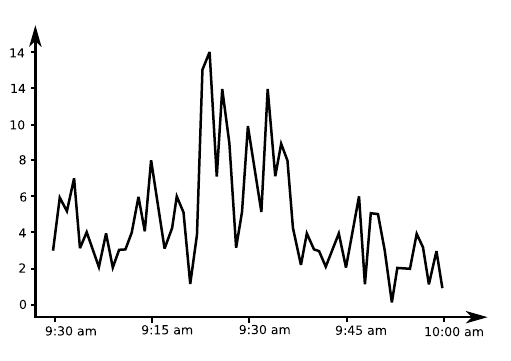}
	\caption{Example of maximum of connections.}
	\label{typical_max}
\end{figure}

\begin{figure}[!t]
	\centering
	\includegraphics[width=2.7in]{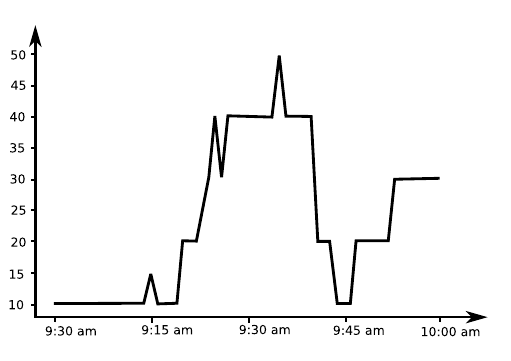}
	\caption{Example of average of connections.}
	\label{typical_avg}
\end{figure}

\begin{figure}[!t]
	\centering
	\includegraphics[width=2.7in]{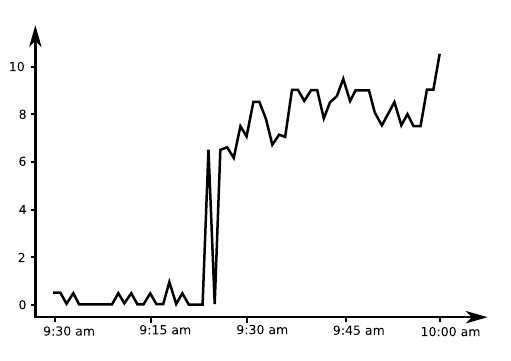}
	\caption{Example of standard deviation of connections.}
	\label{typical_desvest}
\end{figure}

The raw data from AP logs is pre-processed as follows:
\begin{itemize}
	\item The raw data is aggregated by each sampling time-stamp, counting the number of devices per each of the APs.
	\item The resulting table is further aggregated into one-minute basic statistics describing the variability of the device count within each minute. Maximum, minimum, average, standard deviation and variance are calculated over the count values of the samplings performed within the one-minute interval for each AP. Figure \ref{typical_max} shows one example of the behavior of the maximum of connections across an interval of time. Figure \ref{typical_avg} shows the behavior of the average of connections in the same interval. Finally, figure \ref{typical_desvest} shows the behavior of the standard deviation of the number of connections. In the APs from which these examples were chosen, there is a different behaviour roughly around 9:30 am, when a block of classes starts. 
	\item Statistics for all APs are concatenated as a vector, which is the feature vector $x_i$ that will be used as input for the machine learning models.
\end{itemize}

\begin{figure}[!t]
	\centering
	\includegraphics[width=2.7in]{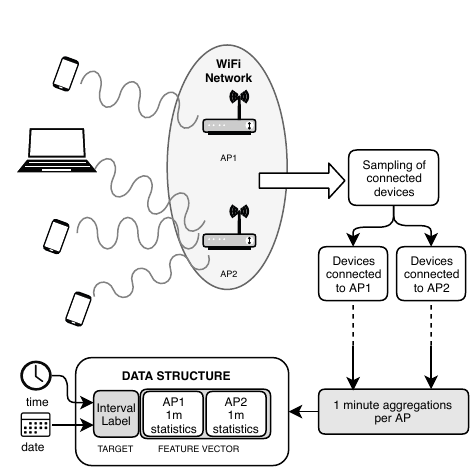}
	\centering	
	\caption{Data pre-processing.}
	\label{dataprocessing}
\end{figure}

Each vector $x_1$ describing the variability of the device counts within the one-minute intervals is assigned a label $y_i$, according to the academic weekly schedule information, depending on the pattern we want to detect in each recognition task. The label  has two possible values, $\mathcal{P}_T$, which indicates the presence of the pattern, and $\mathcal{N}_T$, associated to the absence of the pattern.

\begin{figure}[!t]
	\centering
	\includegraphics[width=2.7in]{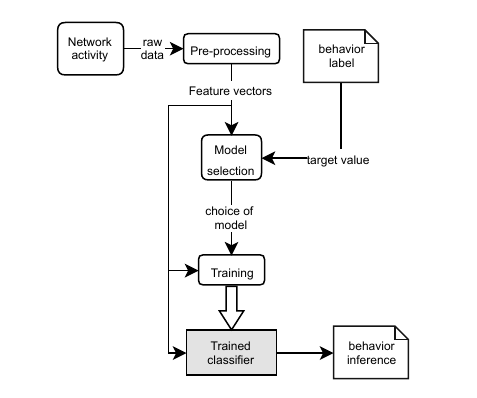}
	\centering	
	\caption{Machine Learning process.}
	\label{mlmodel}
\end{figure}

Various classical machine learning classification techniques are applied, estimating accuracy via 10-fold cross-validation. In a production environment, the fully trained model of choice is then used to perform inference on whether the input shown to each classifier exhibits the pattern corresponding to the label. This implementation would allow the detection of the end of a class, or the detection of a break between classes, without consulting the academic schedule, solely based on the behaviour of the device counts (Fig. \ref{mlmodel}). In all our experiments, a majority of the data was labelled as $\mathcal{N}_T$, producing a situation of class imbalance. We sub-sampled the larger dataset (randomly discarding from each experiment data from the set $\mathcal{N}_T$) to deal with this problem.

The classifiers considered in our experiments were the following: Decision trees, Logistic Regression, Linear Support Vector Machine (SVM), Quadratic SVM, Cubic SVM, Gaussian SVM, K-Nearest Neighbours (kNN), Cosine kNN, Cubic kNN, Weighted kNN, Boosted Trees, Bagged Trees, Subspace Discriminant, Subspace kNN and RUS Boosted Trees

The data was pre-processed in Python with MySQL and the implementation used for the ML methods was that of the Classification Learner App of Matlab R2017b.

\subsubsection*{Labelling process}
One advantage of aiming at recognizing patterns related to the academic schedule, is that it allows the labelling of data without spending time monitoring the behavior of people \emph{in situ} or via video-surveillance as is done in \cite{poucin18}. Instead we use a script that checks for some conditions to determine if the pattern is present or absent and assign the corresponding label, based on the time-stamp of the interval from which $x_i$ was calculated (Fig. \ref{dataprocessing}).

Let $T$ be the experiment at hand, the label are assigned as follows:

\begin{equation}
	y_i = y(x_i) = 
		\begin{cases}
			\mathcal{P}_T, & \text{if the pattern is present in $x_i$}  \\ 
			\mathcal{N}_T, & \text{otherwise, } 
		\end{cases}
		\label{binary}
\end{equation}

 We defined the way the presence of the pattern depending on whether the pattern is associated to an \emph{event} (considered instantaneous), or to a specific time \emph{interval}. 
In the case of the recognition of an event, if time-stamp of a vector is within a range of the event, the pattern is considered to be \emph{present}. For patterns that takes place during an interval the pattern is considered to be present if the one-minute interval from which $x_i$ has been calculated, overlaps with the interval where the pattern takes place. 

\begin{figure}[!t]
	\centering
	\includegraphics[width=2.7in]{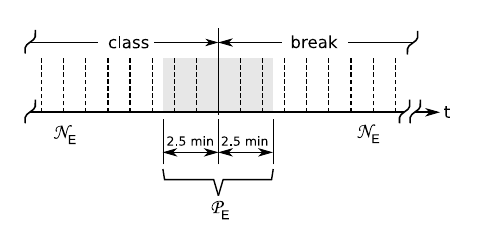}
	\caption{Task 2: Labeling data from the \emph{end of a class}.}
	\label{endofclasslabeling}
\end{figure}

Two binary classification tasks posed::

\begin{itemize}

	\item \emph{Task 1 - Break Interval Recognition:} this problem involves the identification of the 10-minute breaks between class blocks. During this periods, students exit one class to assist to another, or take a break to continue the same class. Some instructors and teachers freely choose to ignore the scheduled break maintaining continuity in the class, or postponing the break to another moment. The labelling strategy is the one defined for patterns that take place in \emph{intervals}. The label $\mathcal{P}_B$ was assigned including a tolerance of one minute before and after the theoretical break interval (as stated in the academic blocks of the schedule) was conceded as tolerance (Fig. \ref{breaklabeling}).
	\item \emph{Task 2 - End of Class Event Recognition:} this problem involves the identification of the movement patterns related to the end of a class block. The labelling strategy used is the one defined for \emph{events}. The label $\mathcal{P}_E$ is assigned within a range of 2.5 minutes from the scheduled end of a class (Fig. \ref{endofclasslabeling}). 

\end{itemize}

as follows:

\begin{figure}[!t]
	\centering
	\includegraphics[width=2.7in]{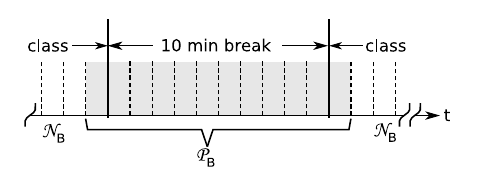}
	\caption{Task 1: Labeling data from a \emph{break interval}.}
	\label{breaklabeling}
\end{figure}

\section{Results and Analysis}

The top performing ML method used in our experiments of binary classification (Bagged Trees)  shows an accuracy (\% of validation data that is well classified) of 93.7 \% for the task of recognizing that a statistics vector belongs to an interval of break between academic blocks, and of 83.3 \% for the task of detecting the event of culmination of a block of classes (Table \ref{results_crossvalidation}).

\begin{table}[]
	\centering	
	\caption{Results for 10-fold cross-validation one week of data.}
	\label{results_crossvalidation}
\begin{tabular}{lcc}
& \% Accuracy &  \% Accuracy \\
ML method  & Break Intervals & End of Class \\
  &  &  \\
Fine Tree			& 83.2 & 64.6 \\
Medium Tree			& 82.8 & 64.6\\
Coarse Tree			& 59.2 & 56.2\\
Logistic Reg		& 70.6 & 45.8\\
Linear SVM			& 72.7 & 59.7\\
Quadratic SVM		& 79.0 & 65.3 \\
Cubic SVM			& 82.4 & 62.5\\
Fine Gaussian SVM	& 76.5 & 56.9\\
Medium Gaussian SVM	& 69.3 & 64.6 \\
Coarse Gaussian SVM	& 55.9 & 50.0\\
Fine kNN			& 83.6 & 70.8\\
Medium kNN			& 73.5 & 52.8\\
Coarse kNN			& 52.1 & 51.4\\
Cosine kNN			& 73.9 & 50.7\\
Cubic kNN			& 70.6 & 49.3\\
Weighted kNN		& 80.3 & 64.6\\
Boosted Trees		& 91.6 & 56.9\\
BaggedTrees			& \textbf{93.7} & \textbf{83.3} \\
Subspace Discriminant	& 76.5 & 54.9 \\
Subspace kNN		& 90.3 & 81.9 \\
RUS Boosted Trees	& 80.7 & 59.7\\

\end{tabular}
\end{table}

The use of Principal Component Analysis was discarded, as its effect in accuracy was not systematically beneficial, ranging from decreases of 18.5 \% to increments of 25.2 \%.

The response of the accuracy to increments of the size of the dataset beyond the data of one week was not beneficial for all the ML techniques. In particular, the Bagging with Decisions trees, as implemented in the Classification Learner App of Matlab R2017b (with 30 learners), saturated on the highest values with the data from one week, which is an indication that further improvement in the classification tasks with the dataset at hand might require noise filtering, better hyper-parameter tuning, more elaborated classification models or more attention to the feature engineering. 

These results show the applicability of  well-known machine learning classification techniques to detect two specific patterns that respond to the structural organization of activities within an academic environment, and should not be generalized to other kinds of buildings, or different types of activities.

In figures \ref{typical_max}, \ref{typical_avg} and \ref{typical_desvest}, there were observable changes in the data around the break interval between 9:20 and 9:30 am. If the activity of people not engaged in the class activities in these intervals of time had been more numerous, or with larger and more active movement patterns, the observable patterns in the data might become invisible within the values generated as a consequence of individuals not engaged in classes at those intervals.

\section{Conclusion and analysis}

We successfully applied machine learning binary classification techniques, on data aggregated via basic one-minute statistics from noisy, incomplete, and imperfect raw data. Particularly stringent conditions were accepted to make the method easily applicable to situations with widespread indoors wireless administrative software. The classification models obtained were able to detect with reasonably high accuracy, common shared mobility patterns in an academic environment. 

Contrary to the general belief stating that there is a hard trade-off between the limitations imposed on machine learning by regulations, and the opportunities to innovate \cite{horvitz15}, the method used open possibilities for further innovation while preserving privacy both in the way data is required, and in the nature of the patterns inferred. Moreover it requires low costs in terms of infrastructure, in comparison to other approaches to gather and analyse data.

Further experiences with strategies like the proposed here, can easily be enriched with information from the context of each environment. There are clear paths of further improvement:

\begin{enumerate}
	\item The temporal nature of the data indicates that the use of time-series models might be promising in more elaborated tasks like forecasting of the connection counts on each AP, and possibly anomaly detection in mobility patterns.
	\item The use of infrastructure, architectural information, as well as the nature of the planned use of the spaces covered by the APs, can also boost our ability to extract knowledge and detect patterns in the behaviour of people.
\end{enumerate}

Obtaining levels of accuracy of 83.3 \% and 93.7 \% in the detection of shared patterns in human mobility behaviour, with a methodology that has innovated in  the sense of pushing down, towards simplicity (not mitigating noise, accepting errors and missing data, and limiting to widespread technology), instead of up (more expensive, complex and scarce technology), demonstrates that the conflict between privacy regulations, budget constraints and the need to innovate in ways to solve our problems, is a promising one.

This proof of concept is only a first step in this research stream. In our future work we will be analysing more complex behaviours, aiming to be able to recognize threats to the security, integrity and health of people, based mainly in WiFi AP connection logs.


\bibliographystyle{IEEEtran}

\bibliography{om_apcount_arxiv}

\end{document}